# Fermi surface dichotomy of the superconducting gap and pseudogap in underdoped pnictides

Y.-M. Xu,[a,b] P. Richard,[c,d] K. Nakayama,[e] T. Kawahara,[e] Y. Sekiba,[e] T. Qian,[c,e] M. Neupane,[a] S. Souma,[d] T. Sato,[e, f] T. Takahashi,[d, e] H.-Q. Luo,[c] H.-H. Wen,[c] G.-F. Chen,[c,g] N.-L. Wang,[c] Z. Wang,[a] Z. Fang,[c] X. Dai,[c] and H. Ding[c, *]

[a]Department of Physics, Boston College, Chestnut Hill, MA 02467, USA

[b]Materials Sciences Division, Lawrence Berkeley National Laboratory, Berkley, CA 94720, USA

[c]Beijing National Laboratory for Condensed Matter Physics, and Institute of Physics, Chinese Academy of Sciences, Beijing 100190, China

[d]WPI Research Center, Advanced Institute for Materials Research, Tohoku University, Sendai 980-8577, Japan

[e]Department of Physics, Tohoku University, Sendai 980-8578, Japan

[f]TRiP, Japan Science and Technology Agency (JST), Kawaguchi 332-0012, Japan

[g]Department of Physics, Renmin University of China, Beijing 100872, China

[*]Electronic address: dingh@aphy.iphy.ac.cn

**Abstract**

High-temperature superconductivity in iron-arsenic materials (pnictides) near an antiferromagnetic phase raises the possibility of spin-fluctuation-mediated pairing. However, the interplay between antiferromagnetic fluctuations and superconductivity remains unclear in the underdoped regime, which is closer to the antiferromagnetic phase. Here we report that the superconducting gap of the underdoped pnictides scales linearly with the transition temperature, and that a distinct pseudogap coexisting with the SC gap develops on underdoping. This pseudogap occurs on Fermi surface sheets connected by the antiferromagnetic wavevector, where the superconducting pairing is stronger as well, suggesting that antiferromagnetic fluctuations drive both the pseudogap and superconductivity. Interestingly, we found that the pseudogap and the spectral lineshape vary with the Fermi surface quasi-nesting conditions in a fashion that shares similarities with the nodal-antinodal dichotomous behaviour observed in underdoped copper oxide superconductors.

**Introduction**

The newly discovered iron arsenic (pnictide) superconductors[1-4] have joined the copper oxide (cuprate) superconductors in the category of high-temperature superconductors (HTCSs). Superconductors are characterized by the size and the symmetry of the superconducting (SC) gap that forms below a critical temperature $T_c$. In optimally hole-doped cuprates, the SC gap function has a *d*-wave symmetry in momentum (*k*) space[5]. Its size evolves smoothly along the Fermi surface (FS) from zero at locations referred to as *nodes* to a large maximum $\Delta$ at the 'antinodes' that puts the value of $2\Delta/k_B T_c$ in the strong coupling regime[6,7]. In contrast, angle-resolved photoemission spectroscopy (ARPES) studies on optimally-hole-doped pnictide $Ba_{0.6}K_{0.4}Fe_2As_2$ ($T_c$ = 37 K; labelled as OPD37K), have revealed isotropic gaps that have different values on different FSs[8-11], with strong pairing occurring on the quasi- (or nearly) nested FSs. The properties associated with the quasi-nested FSs have been described in a previous theoretical work[12].

A remarkable resemblance between these two HTCS systems is the emergence of their SC domes from an antiferromagnetic (AF) phase[13,14] when adding extra carriers (dopants), suggesting that AF fluctuations may have a role in the SC pairing mechanism. For this reason, much attention has been devoted in the past to underdoped (UD) cuprates. In addition to the SC gap, evidence for a large doping-dependent pseudogap (PG) above $T_c$ at the antinodes of UD cuprates has existed for some time[15,16]. Nevertheless, debates on the interplay between superconductivity, antiferromagnetism and PG in cuprates are ongoing. Using the recently discovered pnictides, it is now possible to look into this fundamental issue from a different perspective. Unfortunately, there is no ARPES report on the UD region of hole-doped pnictides investigating the relationship between AF fluctuations and superconductivity. Thus, it is particularly important to conduct a comparison study of optimally doped (OPD) and UD pnictides to determine whether high-temperature superconductivity in cuprates and pnictides can possibly share the same mechanism.

Here we focus mainly on underdoped $Ba_{0.75}K_{0.25}Fe_2As_2$ ($T_c$ = 26 K; labelled as UD26K). We observe that the SC gaps scale with $T_c$ in the underdoped regime, suggesting that the gap size is controlled by the pairing strength. In addition, we reveal the existence of a 18 meV pseudogap on a quasi-nested FS pocket that coexists in momentum with the SC gap at low temperature but survives up to $T^*$ = 115 K- 125 K, temperature at which it completely fills in. Its band-selectivity strongly suggests an origin related to AF fluctuations. Moreover, the evolution of the spectral lineshape on underdoping indicates a dichotomy behaviour between unnested and quasi-nested FSs that shows striking similarities with the nodal-antinodal dichotomy found in cuprates, and that is naturally explained in pnictides in terms of short-range AF scattering.

**Results**

**Superconducting gaps in underdoped pnictides**

As illustrated in Fig. 1d, the FS of UD26K differs from that of the OPD material previously reported[8,9,11,17] only by the size of the holelike and electronlike FS pockets, which become, respectively, smaller and larger in UD26K and satisfy the Luttinger theorem (Supplementary Discussions for the electron counting). Similarly to the OPD sample, the quasi-nesting condition between the inner hole-FS ($\alpha$) and the electron-FSs is maintained, albeit slightly weakened in the underdoped samples. We performed high-resolution measurements on all the FS pockets observed. Below $T_c$, we observe the opening of SC gaps at their corresponding Fermi vectors ($k_F$). As with the OPD compound[8,9,11], these gaps are isotropic within experimental uncertainties (Supplementary Discussions for the superconducting gap symmetry). To characterize these SC gaps further, temperature ($T$)-dependent ARPES measurements have been performed on different FS sheets ($\alpha$, $\beta$ and $\gamma$,$\delta$). We show in Figure 1a-c spectra at various $k_F$ positions, which have been symmetrised to remove the effect of the Fermi function and then normalized by the corresponding normal state spectra. We clearly observe the opening of SC gaps on all the FS sheets at low $T$ and their disappearance above $T_c$. The SC gap value ($\Delta$) on each FS sheet of the UD26K samples is reduced by approximately 30% as compared with the OPD37K samples[8,11], which is almost the same as the reduction of the transition temperature (from 37 K to 26 K). Interestingly, the $2\Delta$ values vary linearly with $T_c$ from moderate underdoping to optimal doping, as illustrated in Figure 1e. Thus, $T_c$ is determined by the energy scale of the pairing gap within this range. The FS-dependent but doping-independent $2\Delta/k_B T_c$ ratio varies from typical BCS values (for the $\beta$ FS) to strong coupling values on the quasi-nested FS sheets. A good agreement between the $T_c$ (doping) dependence of $H_{c2}$ deduced from ARPES and transport measurements supports the bulk nature of the linear scaling of $\Delta$ and $T_c$ (Supplementary Figure S2). It reinforces the notion that $T_c$ in UD pnictides is controlled by the pairing strength, which is apparently different from the trend observed in some studies of cuprates where the SC gap anti-correlates with $T_c$ on underdoping. Our results on UD pnictides are more in-line with a recent report suggesting that the SC gap in UD cuprates at the tip of the nodal 'arc' region scales with $T$c  (ref. 18) and may reflect the 'true' SC gap magnitude.

**Pseudogap and interband scattering in underdoped pnictides**

When looking at the raw spectra recorded above $T_c$ along the $\alpha$ FS of UD26K as shown in Figure 2a, we observe an additional feature around 18 meV with a clear $k$ dependence: whereas a significant decrease of spectral weight below 18 meV is quite obvious in the spectrum recorded along the $\Gamma$-X direction (point no. 1), the spectral weight loss below 18 meV becomes negligible along the $\Gamma$-M direction (point no. 2). Unlike the SC gap, that feature survives in the normal state and hereafter we refer to it as a pseudogap. The temperature-dependence of the PG at point no. 1, given in Figure 2c, is similar to that of UD cuprates: it gradually fills up as temperature increases and eventually closes around $T^* = 115 – 125$ K, a temperature much higher than $T_c$. This observation can be expressed in a more quantitative method using the temperature evolution of the spectral weight loss below 18 meV, which is displayed in Figure 2d. Because below $T_c$ it coexists with the SC gap at a different energy scale, the PG is likely to have an origin that differs from the pairing gap. In fact, the effect of this PG below $T_c$ can be mostly removed by dividing out a spectral function recorded slightly above $T_c$ (45 K), revealing a more pronounced SC coherence peak, as indicated in Figure 1a. However, it is important to stress that unlike the conventional gap-opening phenomenon, there is no long-range magnetic order observed in the UD $Ba_{0.75}K_{0.25}Fe_2As_2$ samples above the SC transition[19-21]. Therefore, the observed PG does not appear to be associated with a phase transition at $T^*$, which is most likely a crossover temperature scale. It is also worth noting that the coexistence of the PG and the SC gap in the same pnictide spectra is similar to recent observation of the coexistence of PG and SC gap at the antinodes of UD cuprates[22].

To understand the origin of the PG, we recorded spectra along the $\beta$ FS as well. In contrast to our observation for the $\alpha$ FS, the spectra along the $\beta$ FS are almost identical, with no sign of PG (see Figure 2e). A natural way to explain the $k$ dependence of the PG is to invoke $(\pi, \pi)$ interband scattering. Indeed, NMR experiments have provided evidence that AF spin fluctuations appear when the holelike FS emerges at the $\Gamma$ point in electron-doped pnictides, thus associating the AF fluctuations with $(\pi, \pi)$ interband scattering[23,24]. As shown in Figure 2f, where the electronlike FS contours centred at the M point are shifted by the $(\pi, \pi)$ wave vector to 'overlap' with the holelike FS contours centred at $\Gamma$, the $\beta$ FS does not significantly overlap with the electronlike FSs. In contrast, we expect much stronger scattering between the $\alpha$ band and the M-centred electronlike bands, especially along the $\Gamma$-X direction. This is consistent with our observation of a well-developed 18 meV energy scale along the $\Gamma$-X direction for the $\alpha$ band of UD26K.

If $(\pi, \pi)$ interband scattering is strong enough to induce a PG in UD pnictides, one expects the quasiparticle (QP) spectral weight to exhibit a similar FS-dependence. In Figure 3a, we compare the spectra of UD26K and OPD37K for different FSs. Whereas the QP spectral weight for the $\beta$ FS remains relatively constant with underdoping, a significant QP weight suppression occurs on the $\alpha$ and the electronlike FS sheets, which are quasi-nested by the $(\pi, \pi)$ wave vector. We can thus summarize our findings on UD pnictides by distinguishing two different behaviours on two types of FS sheets: i) quasi-nested FSs are accompanied by large SC gaps, the formation of a PG, and a strong QP spectral weight suppression on underdoping; ii) FSs that are not quasi-nested are accompanied by a smaller SC gap, the absence of PG, and a quite robust QP on underdoping.

**Discussion**

Even though there exists no direct proof linking the origin of the pseudogap and the pairing mechanism in the cuprates and in the pnictides, it is instructive at this stage to compare the two families of HTCSs. It is well known that UD cuprates also exhibit a *k*-dependent QP weight suppression. Whereas the nodal QP remains robust, the antinodal QP loses integrity quickly with underdoping. This nodal-antinodal dichotomy[25] is illustrated in Figure 3b and 3d for the cuprate superconductor $Bi_2Sr_2CaCu_2O_{8+x}$. In the antinodal region of hole UD cuprates, the QPs can be scattered strongly by spin fluctuations via the $(\pi, \pi)$ AF wave vector. Similarly, the SC phase of the pnictides emerges when doping away from the AF spin-density-wave phase in the parent compounds, and AF fluctuations exist in the UD region[19,24,26]. Such an AF-fluctuations scenario is consistent with the observed FS-dependent QP suppression. On sections of FS that are approximately connected by the AF wave vector, the similarities in the spectral lineshape of pnictides and cuprates are striking. In the overdoped and OPD regions, both materials show a well-defined Bogoliubov quasiparticle peak in the SC state and a broadened single peak above $T_c$, as shown in Figure 3b-c. On underdoping, the SC Bogoliubov quasiparticle weight of both materials is rapidly weakened, and the near $E_F$ spectral function above $T_c$ switches from a peak to a dip.

The origin of the PG is clearly an important part of the physics of both families of HTCSs. In the pnicitides, it is quite natural to attribute its origin to AF fluctuations since it emerges near an AF phase, and its properties are shaped by the AF wave vector. Our observations of the SC coherence suppression and the reduction of the SC gap on underdoping, as well as the emergence of a PG that coexists with the SC gap below $T_c$, support the scenario where the PG is the offspring of a competing state, although the SC pairing itself may be mediated by AF fluctuations, as suggested by various experiments[8,24,26,27] and theoretical works[12,28-31]. In particular, a functional renormalization group approach has been applied to both cuprates and pnictides, suggesting that the pairing in both materials could be driven by AF correlations and enhanced by $(\pi, \pi)$ FS quasi-nesting[32]. We caution that whereas AF correlations increase in pnictides on underdoping from the optimal concentration[24], the FS quasi-nesting conditions degrade slowly. This possibly implies that long-range spin-density-wave in the pnictides is not driven solely by FS nesting, as previously suggested theoretically[33]. Nevertheless, the stronger SC pairing ($2\Delta/k_B T_c \sim 7$) and the PG observed in the quasi-nested FS sheets connected by the AF wave vector support the AF origin of both the SC gap and the PG in the pnictides. The observation of similar dichotomous behaviours in UD pnictides and UD cuprates raises the possibility for a unifying picture of HTCSs involving AF fluctuations and calls for further experimental and theoretical works beyond the scope of the present study.

**Methods**

**Samples preparation**

The high quality single crystals of $Ba_{0.75}K_{0.25}Fe_2As_2$ used in our study were grown by the self-flux method. The $T_c$ of the sample has been estimated to be $T_c^{mid}$ = 26 K from magnetic susceptibility measurements, with a transition width of about 3 K (Supplementary Figure S1).  The nominal composition is consistent to the K content determined by energy-dispersive X-ray spectroscopy. Low-energy electron diffraction on a measured surface shows a sharp 1x1 patterns indicating that the sample surface has a fourfold symmetry without any detectable surface reconstruction down to 80 K.

**ARPES experiments**

High-resolution ARPES measurements were performed in the photoemission laboratory of Tohoku University using a high-flux microwave-driven Helium source (He I$\alpha$ resonance line, $h\nu$ = 21.218 eV) with an energy resolution of 3 meV, and a momentum resolution of $0.007Å^{-1}$. We used a VG-SCIENTA ESE2002 spectrometer as detector. Samples were cleaved *in situ* at 10 K and measured at 5 – 50 K in a working vacuum better than $5x10^{-11}$ Torr (for the PG measurements, the temperature was increased up to 125 K and the vacuum remained better than $1x10^{-10}$ Torr at 125 K). Some experiments were also performed at the Wadsworth beamline of Synchrotron Radiation Center with an energy resolution of 10 meV and a momentum resolution of 0.02 $\text{Å}^{-1}$. A VG-SCIENTA R4000 analyzer was used for synchrotron measurements. The mirror-like sample surfaces were found to be stable without obvious degradation during measurement periods of 3 days (30 hours for synchrotron measurements). The Fermi level of the samples was referenced to that of a gold film evaporated onto the sample holder. We observed that the SC gap closes around the bulk $T_c$ (see Figure 1), which suggests that the present ARPES results reflect the bulk properties. We have confirmed the reproducibility of the ARPES data on more than ten sample surfaces and through thermal cycles across $T_c$.

## Acknowledgements

We thank J. H. Bowen for proofreading our manuscript. We acknowledge the support through grants from the Chinese Academy of Sciences, NSF, Ministry of Science and Technology of China, TRiP-JST, CREST-JST, JSPS and MEXT of Japan, and NSF, DOE of US. This work was based on research conducted at the Synchrotron Radiation Center supported by NSF No. DMR-0537588.

## Author Contributions

Y.-M.X., P.R., K.N., T.K., Y.S., T.Q., M.N., S.S., and T.S. performed experiments, Y.-M.X., P.R., and K.N. analyzed data, H.D. and Y.-M.X designed experiments, H.D., Y.-M.X. P.R., T.S. T.T., Z.W., Z.F., X.D. wrote the paper, H..-Q.L., H.-H.W., G.-F.C., N.-L.W synthesized materials. All authors discussed the results and commented on the manuscript.

## Competing Financial Interests

We declare no competing financial interests.

## Figure Legends

Figure 1. (colour)

**SC gaps and FS of UD pnictides.**

Panels (**a**-**c**) show the $T$-dependence of the symmetrised energy distribution curves (EDCs) of $Ba_{0.75}K_{0.25}Fe_2As_2$ ($T_c$ = 26 K) divided by a smoothed spectrum recorded above $T_c$ (45 K – 50 K) on the $\alpha$, $\beta$ and $\gamma,\delta$ FSs, respectively. The corresponding momentum locations are given in the inset of (**c**) (red, blue and green dots for the $\alpha$, $\beta$ and $\gamma,\delta$ FSs, respectively). Dashed lines indicate the QP peak position at low $T$, which can be regarded as the SC gap value. We find $\Delta_\alpha(0)$ = 8.5 meV, $\Delta_\beta(0)$ = 4.0 meV and $\Delta_{\gamma,\delta}(0)$ = 7.8 meV. The raw curves obtained at 10 K (blue) and 45 K (red) for the $\alpha$ FS, as well as the smoothed 45 K (black dashed), are shown at the bottom of panel (**a**) as examples. (**d**) Comparison of FS contours between the OPD37K and UD26K samples. The dots indicate Fermi wave vectors extracted from ARPES measurements above $T_c$ by tracing the peak position of momentum distribution curves (MDCs). (**e**) Full SC gap values (2$\Delta$) versus $T_c$. The dashed red, blue and green lines are linear fits of the SC gaps on the $\alpha$, $\beta$ and $\gamma,\delta$ FSs passing through the origin. The dashed black line is the BCS line with a slope of 3.52$k_B$. We defined the vertical error bars as twice the derivation that we get by fitting the SC coherent peaks with Lorentzien functions. The horizontal error bars represent the transition width measured by transport experiments.

Figure 2. (colour)

**Momentum dependence of the PG in Ba$_{0.75}$K$_{0.25}$Fe$_2$As$_2$ ($T_c$ = 26 K).**

(**a**) Symmetrised EDCs of UD26K pnictide measured below and above $T_c$ at two different locations on the $\alpha$ FS (points #1 and #2 in panel (**b**)). The red arrow indicates that the PG is ~18 meV, and the dashed vertical line shows that the SC gap on the $\alpha$ FS is ~8 meV. (**b**) Schematic FS plot near the $\Gamma$ point indicating the measurement locations of spectra presented in (**a**), (**c**) and (**e**). (**c**) $T$- dependence of the symmetrised EDCs of UD26K measured at point #1 on the $\alpha$ FS above $T_c$. The vertical dashed line indicates the energy scale of the PG (18 meV). The shaded regions represent the spectral weight loss in the PG state. It is obtained by subtracting the symmetrised EDCs from a quadratic background. (**d**) $T$-dependence of the relative weight loss (normalized by the one obtained at $T = 45$ K) in the PG state of UD26K. The error bars represent the uncertainty in calculating the relative weight loss. (**e**) Similar as (**a**) but for the $\beta$ FS (points #3, #4 and #5). The dashed line shows the SC gap on the $\beta$ FS (4 meV). (**f**) Electronlike FS contours of UD26K shifted to the $\Gamma$ point by the $(\pi, \pi)$ wave vector to illustrate the quasi-nesting conditions. The regions where the electronlike and holelike FSs 'overlap' are highlighted by shadowed areas.

Figure 3. (colour)

**Comparison of the dichotomous behaviour in the spectral lineshape of UD cuprates and UD pnictides.**

(**a**) Comparison between the EDCs of OPD37K (dashed line) and UD26K (solid line) samples below $T_c$ on different FS sheets (red, green, and blue dots in the inset). (**b**) Comparison of the symmetrised EDCs at the antinodal $k_F$ point of an overdoped (OD) Bi2212 sample ($T_c = 82$ K; dashed lines) and an UD Bi2212 sample ($T_c = 83$ K; solid lines), measured below (blue) and above (red) $T_c$. The black dashed (doted) line indicates the zero intensity position of the OD Bi2212 (UD Bi2212) sample. (**c**) Comparison between the symmetrised EDCs of OPD37K (dashed) and UD26K (solid) at the $k_F$ point of the $\alpha$ FS along $\Gamma$-M. Blue and red curves refer to spectra measured below and above $T_c$, respectively. (**d**) Comparison between the EDCs of the OPD (red) and UD (black) $Bi_2Sr_2CaCu_2O_{8+x}$ (Bi2212) samples measured at 10 K at the nodal point. (**e**) Schematic diagrams illustrating that the AF $(\pi, \pi)$ wave vector connects the antinodal regions in hole-doped cuprates, and the inner holelike FS and the electronlike FSs in hole-doped pnictides.

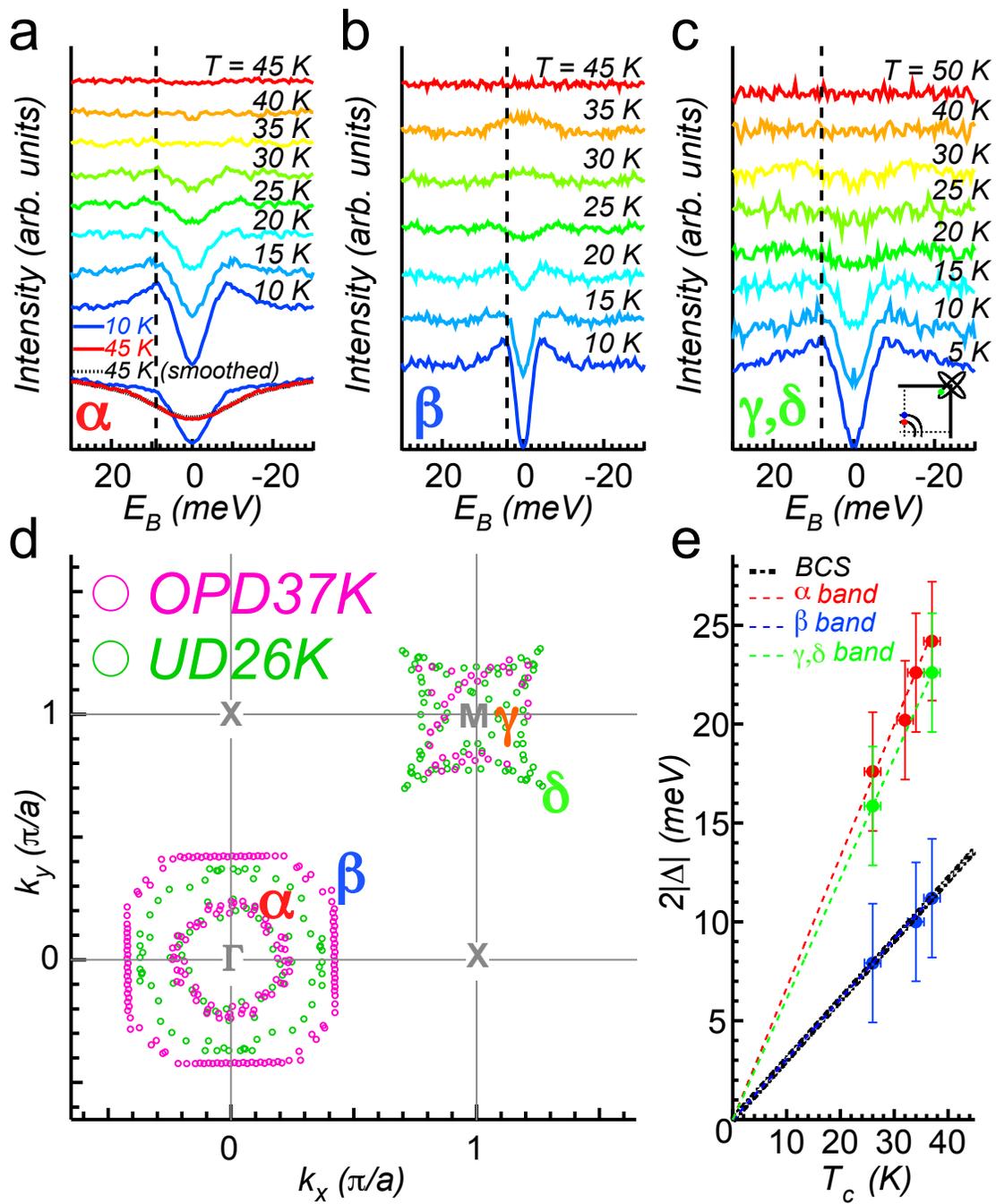

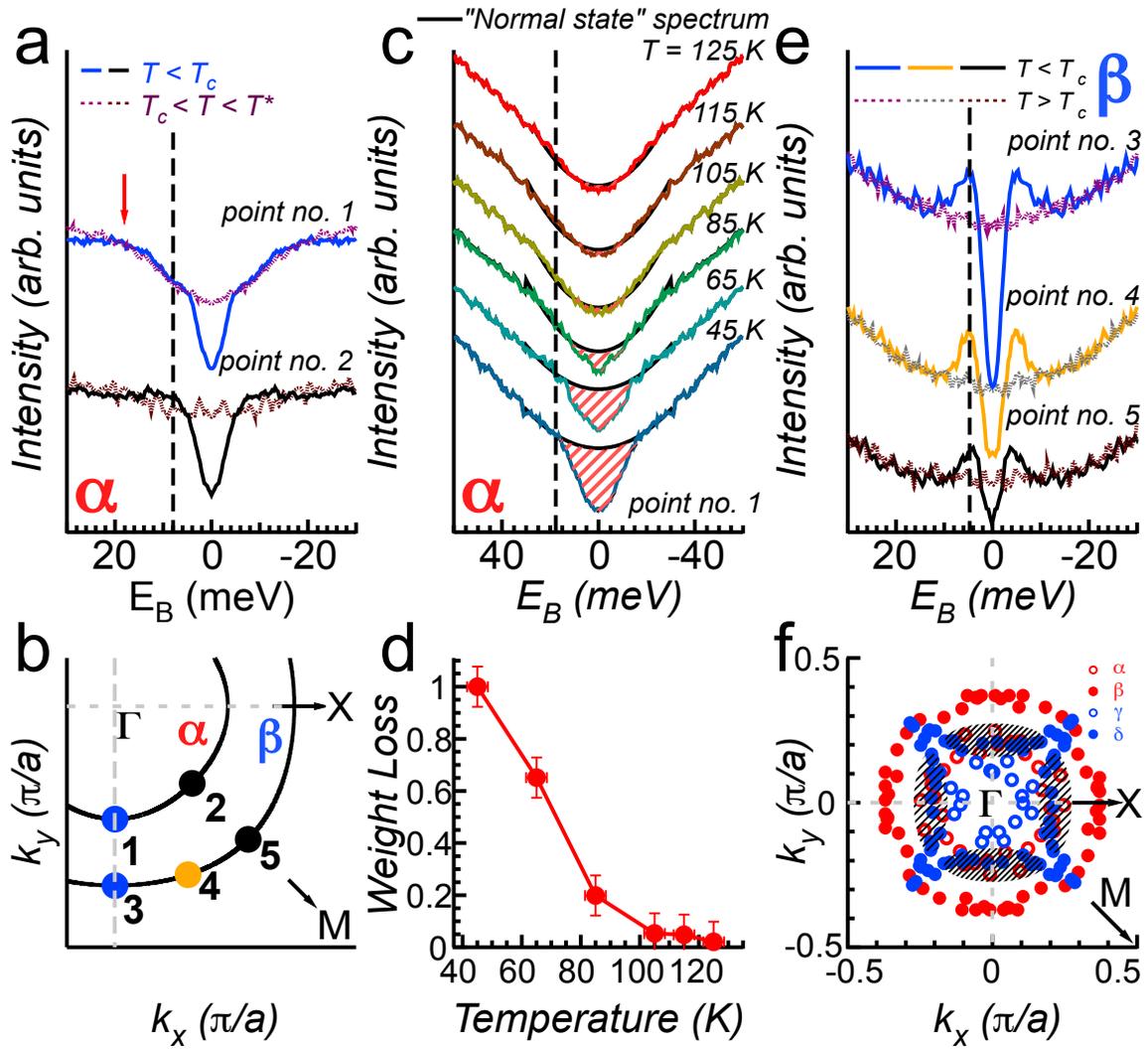

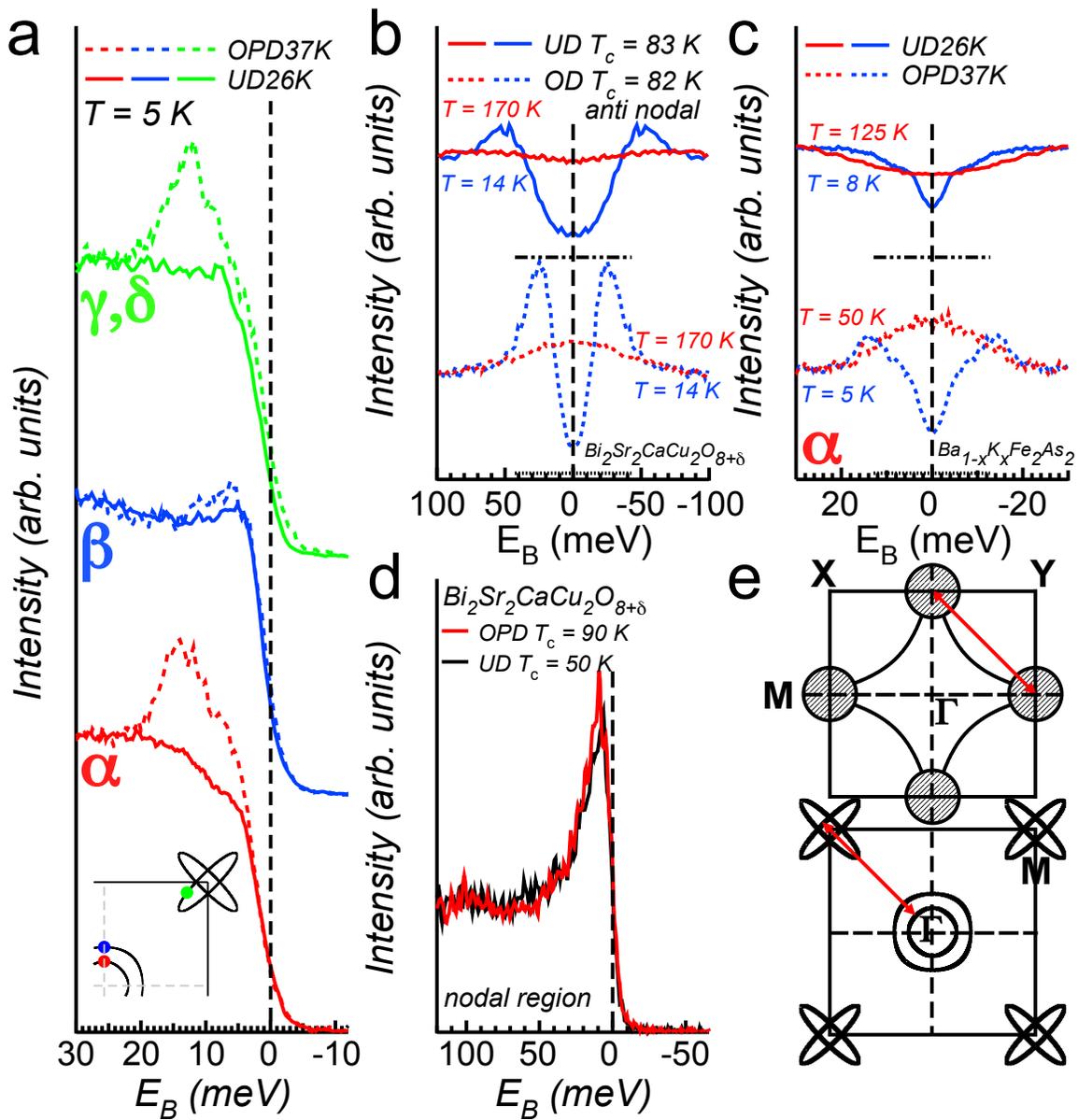

**Supplementary Figures**

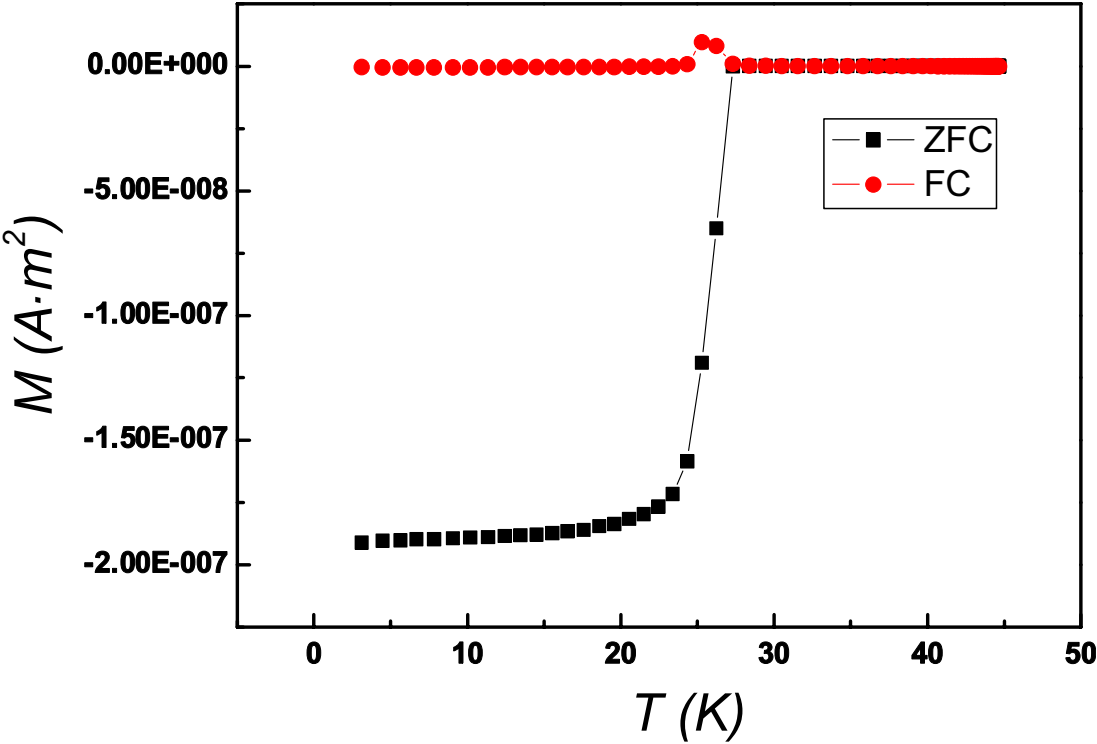

**Supplementary Figure S1.** (colour) Magnetic susceptibility of $Ba_{0.75}K_{0.25}Fe_2As_2$ as function of $T$ ($T_c$ = 26 K).

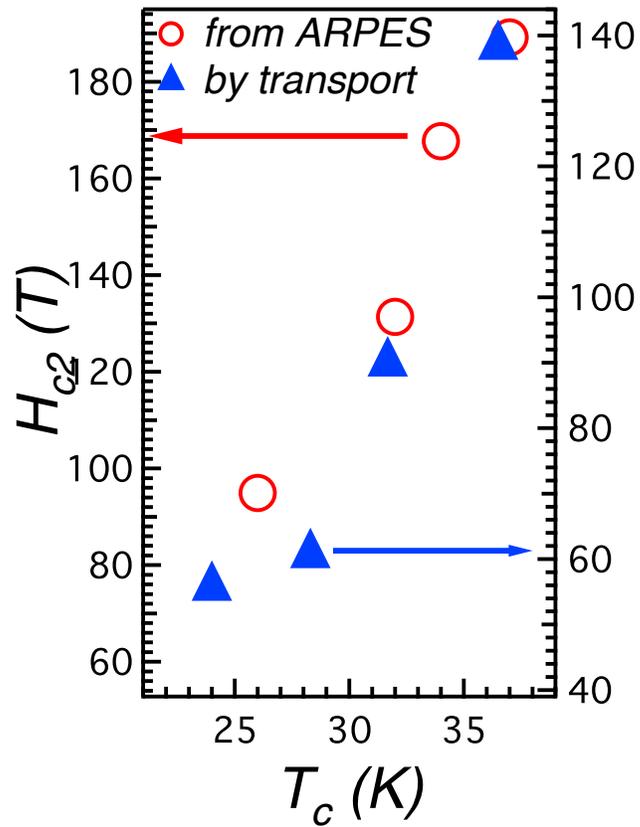

**Supplementary Figure S2. Doping dependence of $H_{c2}$.** (colour) $T_c$ dependence of $H_{c2}$. Blue triangles are obtained from transport measurements. Red circles are extracted from the ARPES measurements of the gap value and the Fermi velocity via the BCS relation described above.

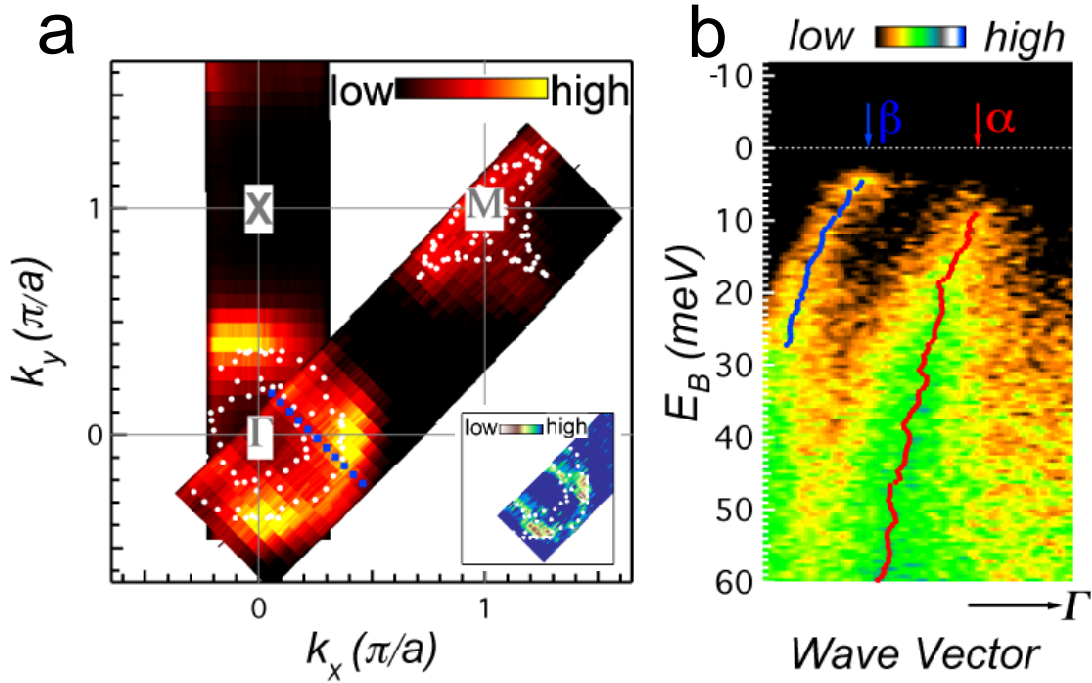

**Supplementary Figure S3. Fermi surface and band dispersions of Ba$_{0.75}$K$_{0.25}$Fe$_2$As$_2$ ($T_c$ = 26 K).** (colour) (a) ARPES spectral intensity plot integrated within ±10 meV of $E_F$ as function of momentum indicating the FS contours. White dots representing the Fermi wave vectors are extracted from the peak position of the MDCs above $T_c$. The inset shows the ARPES second derivative spectral intensity around Γ. (b) ARPES spectral intensity plot measured at 5 K as a function of momentum and binding energy along the cut in momentum space indicated by the blue dashed line in panel (a). Two holelike (α and β) band dispersions are well evidenced. Red (blue) dots are the extracted band dispersion of the α (β) band determined from the MDC peak positions.

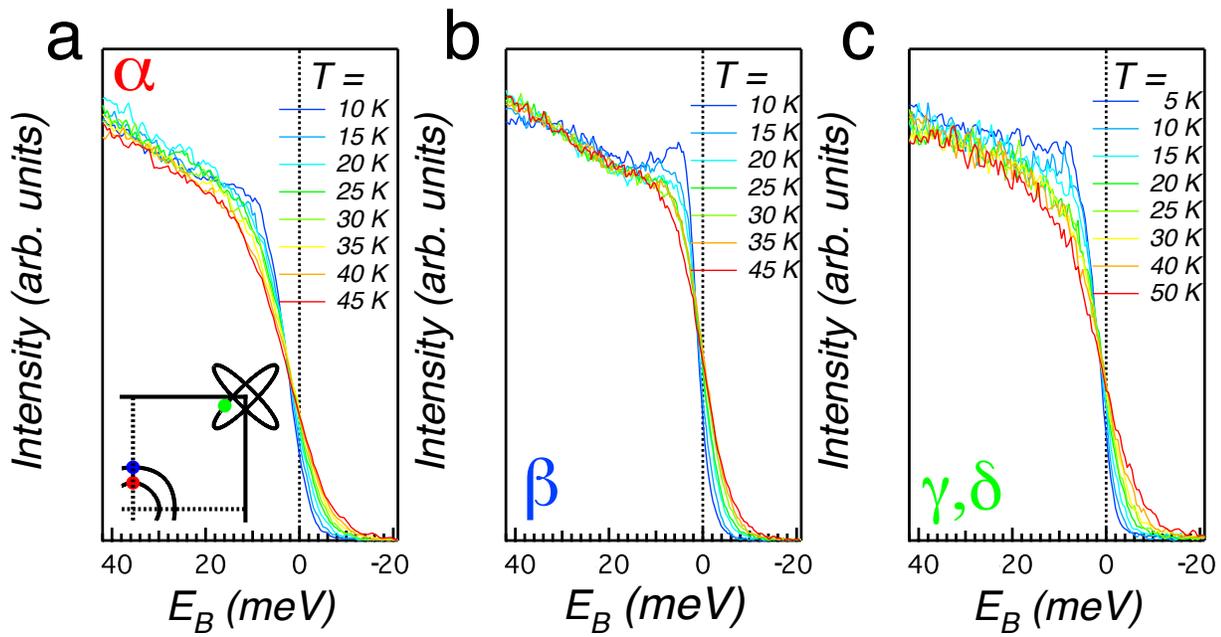

**Supplementary Figure S4. Superconducting gap symmetry in EDCs of Ba$_{0.75}$K$_{0.25}$Fe$_2$As$_2$ ($T_c$ = 26 K).** (colour) $T$-dependence of EDCs measured on (a) the $\alpha$ FS, (b) the $\beta$ FS and (c) the $\gamma$,$\delta$ FS. The inset in (a) gives the momentum location.

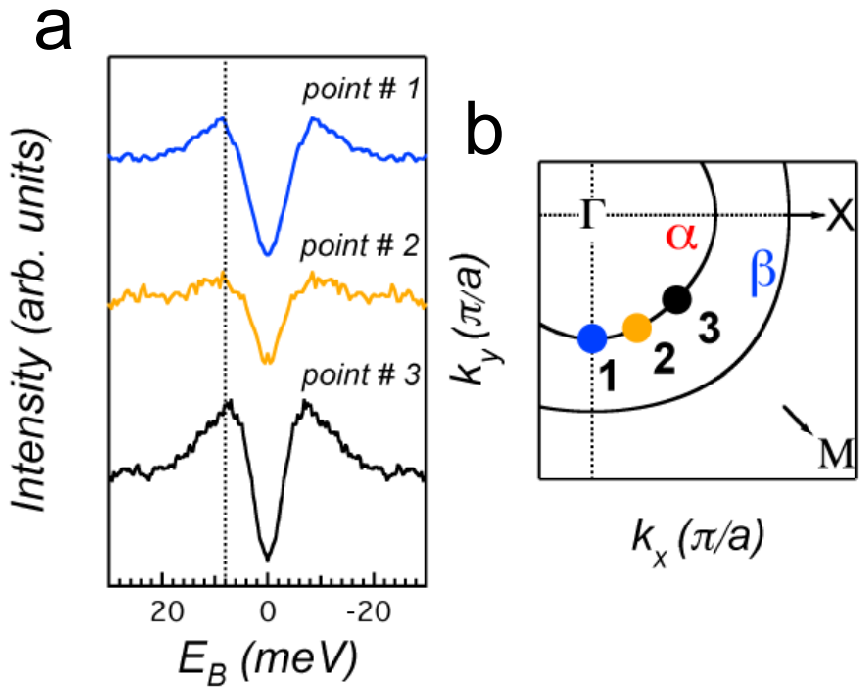

**Supplementary Figure S5. Nearly isotropic superconducting gap on the α FS of Ba$_{0.75}$K$_{0.25}$Fe$_2$As$_2$ ($T_c$ = 26 K).** (colour) (a) Symmetrised EDCs at low temperatures (<10 K) divided by a corresponding smoothed normal-state spectrum measured at three different *k* locations on the α FS; (b) Schematic FS plot indicating the measurement locations in panel (a).

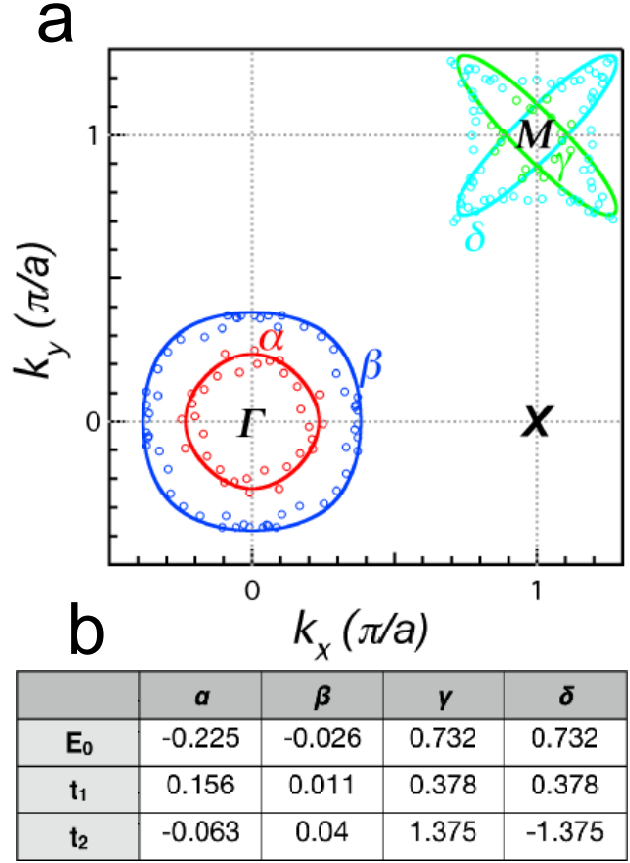

**Supplementary Figure S6. Tight-binding fitting results of Ba$_{0.75}$K$_{0.25}$Fe$_2$As$_2$ ($T_c$ = 26 K).** (colour) (a) Measured FS (circles) and tight-binding fitting curves (solid lines); (b) Tight-binding parameters (in the unit of eV) obtained from the following formula[34]:

$$E^{\alpha,\beta}\left(k_x,k_y\right) = E_0^{\alpha,\beta} + t_1^{\alpha,\beta}\left(\cos k_x + \cos k_y\right) + t_2^{\alpha,\beta}\cos k_x \cos k_y$$

$$E^{\gamma,\delta}\left(k_x,k_y\right) = E_0^{\gamma,\delta} + t_1^{\gamma,\delta}\left(\cos k_x + \cos k_y\right) + t_2^{\gamma,\delta}\left(\cos \frac{k_x}{2}\right)\left(\cos \frac{k_y}{2}\right).$$

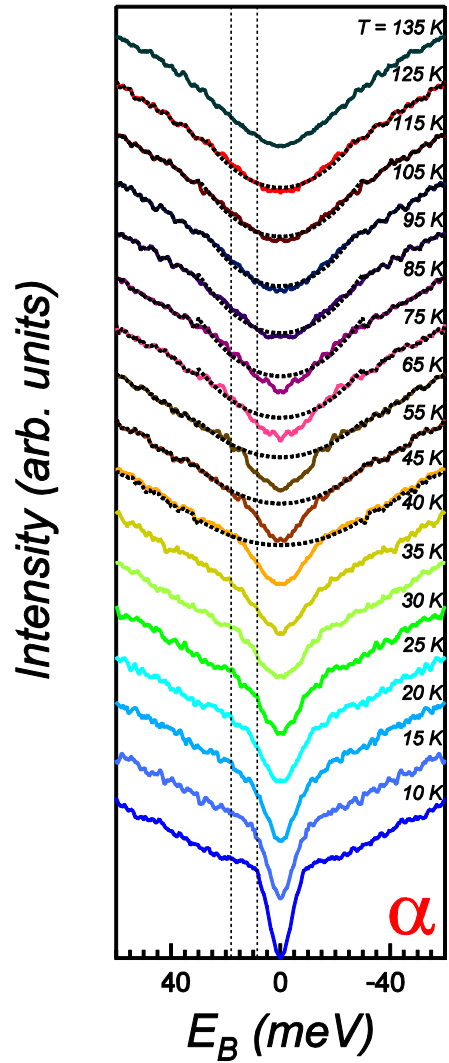

**Supplementary Figure S7. Temperature dependence of symmetrised EDCs on the α FS of Ba$_{0.75}$K$_{0.25}$Fe$_2$As$_2$ ($T_c$ = 26 K).** (colour) $T$-dependence of symmetrised EDCs measured on the α FS. The dashed vertical lines show that the SC gap and the pseudogap on the α FS are ~8 meV and ~18 meV, respectively. The corresponding quadratic background curves for $T$ = 45 − 125 K are shown in dashed line next to the symmetrised EDCs.

## Supplementary Discussion

### Magnetic susceptibility of $Ba_{0.75}K_{0.25}Fe_2As_2$ ($T_c$ = 26 K)

The magnetic susceptibility of UD26K indicates a critical temperature of 26 K, defined at the midpoint of the transition, and a transition width of about 3 K (from 10% to 90 %) as shown in Supplementary Figure S1. Such a transition width is comparable to that of OPD37K samples.

### Doping dependence of $H_{c2}$

Since ARPES is a surface sensitive technique, it is important to compare the energy scale of the SC gap to that inferred from independent bulk measurements. For this purpose, we calculated the values of $H_{c2}$ for various doping levels from the BCS relation using quantities obtained by ARPES. In the clean limit, $H_{c2}$ is related to $\Delta$ by the BCS formula

$$H_{c2} = \Phi_0 \left/ \left[ 2\pi \left( \frac{\hbar v_F}{\pi \Delta} \right)^2 \right] \right., \text{ where } \Phi_0 = 2.07 \times 10^{-15} Wb \text{ is the flux quantum and } v_F \text{ is the Fermi}$$

velocity. For simplicity, we used the values of $v_F$ and $\Delta$ measured by ARPES for the $\alpha$ FS sheet, which has the largest pairing gap (the value of $v_F$ is ~0.5 eV·Å [7.6 x $10^4$ m/s] in UD26K, which is similar to the one obtained in OPD37K[17]). As shown in Supplementary Figure S2, the extracted values are consistent within a factor smaller than 2 with the $H_{c2}$ values obtained from transport measurements, which is quite good considering the single-band approximation used here and the fact that the measured $H_{c2}$ itself has a sizable uncertainty due to the high-field extrapolation.

### Fermi surface of $Ba_{0.75}K_{0.25}Fe_2As_2$ ($T_c$ = 26 K) and electron counting

As compared to the optimally doped compound, the areas of the two holelike FSs centred at the $\Gamma$ point shrink from ~4% and ~18% to ~4% and ~13% of the Brillouin zone (BZ), while the areas of the electronlike FSs centred at the M point increase from ~2% and ~4% to ~2% and ~6% of the BZ, respectively. According to Luttinger's theorem and assuming a double degeneracy for the inner hole ($\alpha$) FS[8,11], the implied hole concentrations are ~40% and ~26% hole/2Fe for the OPD37K and UD26K samples, respectively, in good agreement with their K contents.

**Superconducting gap symmetry in EDCs of $Ba_{0.75}K_{0.25}Fe_2As_2$ ($T_c$ = 26 K)**

Supplementary Figure S4 shows the temperature evolution of EDCs located on various FSs. To extract the values of the SC gap, we followed a common practice in ARPES and symmetrised the EDCs at $k_F$ to approximately remove the effect of the Fermi-Dirac function. In addition, we divided each symmetrised EDC by the corresponding normal-state spectrum (45 – 50 K) to remove the influence of the V-shaped spectral background. Figs 1a-c in the main text show the results obtained for the $\alpha$, $\beta$ and $\gamma,\delta$ FS, respectively. A small but discernable coherence peak emerges at low $T$ (5 – 10 K) at a binding energy of ~8.5 meV, ~4.0 meV and ~7.8 meV for the $\alpha$, $\beta$ and $\gamma,\delta$ FS, respectively, which can be regarded as estimated values of the SC gaps.

We investigated further the properties of the SC gap by measuring its momentum ($k$)-dependence on the $\alpha$ and the $\beta$ FS sheets. In Supplementary Figure S5 and Fig. 2e in the main text, we show symmetrised spectra below $T_c$ for 3 different locations on the $\alpha$ FS and the $\beta$ FS. We find no significant variation of the SC gap size around both FSs (dashed curves in both panels) at 21.218 eV, suggesting that the nearly isotropic symmetry of the SC gaps observed in OPD37K samples[8,11] is maintained in the moderately UD region. We caution that a small modulation of the SC gap along the c axis, which may also be present in UD26K, has been reported for the $\alpha$ band of OPD37K[35].

**Temperature dependence of symmetrised EDCs on the $\alpha$ FS of Ba$_{0.75}$K$_{0.25}$Fe$_2$As$_2$ ($T_c$ = 26 K)**

As shown in Supplementary Figure S7, full temperature cycle (10 – 135 K) symmetrised EDCs indicate that the superconducting gap (~8 meV) exists with a pseudogap (~18 meV) below $T_c$, but disappears above $T_c$ while the pseudogap continuously fills up to $T^*$ (between 115 K and 125 K).